\documentclass[preprint,final]{aastex}
\shortauthors{Nataf et al.}
\usepackage{subfigure}
\shorttitle{Intrinsic Magnitude Dispersion}
\usepackage{natbib}
\usepackage{float}
\usepackage{sidecap}

\begin{document}

\title{The Intrinsic I-Band Magnitude Dispersion of the Galactic Bulge Red Clump }
\author{D. M. Nataf\altaffilmark{1}, A.Udalski\altaffilmark{2}}
\altaffiltext{1}{Department of Astronomy, Ohio State University, 140 W. 18th Ave., Columbus, OH 43210}
\altaffiltext{2}{Warsaw University Observatory, Al. Ujazdowskie 4, 00-478 Warszawa,Poland}
\email{nataf@astronomy.ohio-state.edu}

\begin{abstract}
We measure the intrinsic magnitude dispersion of the Galactic bulge red clump (RC) using two different methods and arrive at an estimate of $\sigma_{I} \approx 0.17$ mag. We first estimate the width of the RC by analyzing in detail a sample of RC measurements toward double RC sightlines, which are toward regions of the bulge that are geometrically thin. We obtain $\sigma_{I} \sim 0.17$ mag. We then estimate the value by comparing the luminosity function for bulge RC and red giant (RG) stars in $V$ to that in $I$ for a sightline near the plane, and we obtain a value of $\sigma_{I} \sim 0.14-0.17$ mag. This result has structural and evolutionary repercussions. It  constrains model estimates of the bar's orientation angle derived from RC studies as well as the inferred characteristic length of the bar for a given characteristic height. Moreover, the value is too large to account for using the predicted evolutionary effects of the known metallicity distribution, implying the presence of other population effects.
\end{abstract}

\keywords{Galaxy: bulge}

\section{Introduction}
\label{sec:Introduction}
The red clump (RC) of the Galactic bulge has a vibrant history as an investigative tool to enable study of the bulge stellar population, its age, its composition, its distribution, and its dynamics. Its number counts relative to the red giant branch have been used to either infer a significant intermediate-age population \citep{1994AJ....107.2060P} or modestly enhanced helium enrichment \citep{1994A&A...285L...5R,1995A&A...300..109M} for the bulge. \citet{1994ApJ...429L..73S,1997ApJ...477..163S} used OGLE observations of the variation of the dereddened RC mean brightness, brightness dispersion and number counts with longitude and latitude to infer that the bulge stars were distributed as bar, with the near side oriented toward positive longitudes. The angle between the bar's major axis and the line of sight between the Sun and the Galactic center, denoted $\alpha$, was constrained to the range $20^{\circ} \leq \alpha \leq 30^{\circ}$. \citet{2007MNRAS.378.1064R} obtained similar results using the $\sim 11 \times$ larger OGLE-II survey.

Observations of the RC have also been used to trace the reddening toward the bulge \citep{2003ApJ...590..284U,2004MNRAS.349..193S,2009ApJ...696.1407N}, to study its dynamics \citep{2007MNRAS.378.1165R,2010A&A...519A..77B,2011arXiv1104.0223D}, and to model the geometry of the bar at large longitudinal separations from the Galactic center \citep{2005ApJ...630L.149B,2007AJ....133..154L,2007A&A...465..825C,2008A&A...491..781C}. It has recently been discovered that the RC is split into two components for sightlines near the bulge minor axis and at least 5 degrees ($\sim$700 pc) removed from the plane \citep{2010ApJ...721L..28N,2010ApJ...724.1491M}. This effect is likely due to the Milky Way's bulge being X-shaped at large separations from the plane, a feature predicted by N-body models to be ubiquitous in bars \citep{2005MNRAS.358.1477A} and frequently observed in the bars of other galaxies \citep{2006MNRAS.370..753B}.

What allows the RC to be such a powerful diagnostic tool is that it occupies a very narrow range in color-magnitude space with very limited population effects. This is both expected from theory \citep{2001MNRAS.323..109G,2002MNRAS.337..332S}, observed in the local Hipparcos population \citep{2000ApJ...531L..25U,2008A&A...488..935G}, and in the stellar populations of other galaxies \citep{2010AJ....140.1038P}.

However, while the intrinsic magnitude dispersion of the RC is indeed smaller than that of other stellar populations, its exact size is not known. In their study of the structure of the bar, \citet{1997ApJ...477..163S} showed that the axis ratios and orientation angle of the bar are degenerate with the assumed value of the intrinsic dispersion. This is due to the fact that the observed brightness dispersion is a quadratic sum of the intrinsic dispersion and the geometric dispersion (stars at different distances will have different apparent magnitudes), with the geometric dispersion being a strong function of the bar structure. A larger assumed value for intrinsic dispersion, and thus a lower inferred value for the geometric thickness of the bar, pushes the models to smaller values of the axis ratio $x_{0}$:$z_{0}$ (a decreased characteristic length for the bar along its major-axis relative to its characteristic height) as well as a larger value for $\alpha$. In this Letter, we estimate the value of the intrinsic dispersion of the Galactic bulge RC, that is, the dispersion that would be measured if all of the stars were at the same distance and if there was no differential reddening.  

We infer the value of the intrinsic magnitude dispersion using two independent methods and estimate it to be $\sigma_{I} \approx 0.17$ mag. With our first method, we measure the brightness dispersion of RC stars in the X-wings of the Milky Way bulge using the results of \citet{2010ApJ...721L..28N}. As the X-wings of the bulge are expected from both observations and dynamical models to be substantially thinner than the underlying ellipsoid-like structure, the observed dispersion should correspond closely to the intrinsic dispersion. With our second method, we compare the observed brightness dispersion for the RC in $V$ and $I$ using clump-centric color-magnitude diagrams (CCCMD), a diagnostic tool to investigate large stellar populations with variable reddening and geometry that is introduced and described in \citet{2011ApJ...730..118N}. 

The structure of this Letter is as follows. We briefly summarize the data used in Section \ref{sec:Data}. The contributing factors to the intrinsic dispersion are estimated from theory in Section \ref{sec:sources}. A purely geometrical estimate using observations of the double clump are summarized in Section \ref{sec:DoubleClump}, whereas the method of the CCCMD and the results derived thereof are presented in Section \ref{sec:CCCMD}.  We discuss our findings and their implications in Section \ref{sec:Conclusion}.

\section{Data}
\label{sec:Data}
OGLE-III observations were obtained from the 1.3 meter Warsaw Telescope, located at the Las Campanas Observatory in Chile. Detailed descriptions of the instrumentation, photometric reductions and astrometric calibrations are available in \citet{2003AcA....53..291U}, and \citet{2008AcA....58...69U}.

\section{The Sources of Intrinsic Brightness Dispersion}
\label{sec:sources}

The main contributing factors to the intrinsic dispersion of the bulge RC are expected to be the evolution of RC stars during their $\sim$100 Myr lifetime, the effects of dispersion in [Fe/H], and that of the dispersion in [$\alpha$/Fe]. Other factors such as photometric noise, blending due to binary contamination, and differential reddening  should be no more than a few hundredths of a mag each, and will thus not significantly contribute in quadrature. Age is expected to be a very marginal factor: \citet{2001MNRAS.323..109G} estimate that for a $\sim$10 Gyr old, solar-metallicity population, $dM_{V}/dt$ and $dM_{I}/dt$ will both be $\sim$0.02 mag Gyr$^{-1}$. We note that \citet{2001MNRAS.323..109G} predicted an intrinsic dispersion of 0.051-0.107 mag for the bulge RC in $I$ using the metallicity constraints available at the time. We refer the reader to their section 5.1. We perform a similar procedure here for illustrative purposes, as well as to employ the benefit of more recent data constraints.

The brightness dispersion a single population would have is estimated using HST observations of 47 Tuc \citep{2007AJ....133.1658S}. 47 Tuc is likely to be a single-age, single-metallicity population with a small spread ($\sim$3\%) in initial helium abundance \citep{2010MNRAS.408..999D,2011arXiv1102.3916N}. We draw a box around the RC as in \citet{2011arXiv1102.3916N} and find a brightness dispersion of 0.065 mag in $I$ and 0.061 mag in $V$. This value includes the evolution of the RC during its $\sim$100 Myr lifetime as well as the variation induced by any possible stochastic mass loss  during the first ascent of the red giant (RG) branch. 

Metallicity dispersion will be a contributing factor to the brightness dispersion. \citet{2001MNRAS.323..109G} predict that for a 10 Gyr population near solar metallicity, $dM_{V}/d\rm{[Fe/H]}$ will be 0.60 mag dex$^{-1}$ and $dM_{I}/d\rm{[Fe/H]}$ will be 0.21 mag dex$^{-1}$, with Iron-rich stars being \textit{fainter}. The effect of 0.25 dex of enhancement in [$\alpha$/Fe] is predicted to be 0.15 mag in $V$ and 0.07 mag in $I$, with $\alpha$-enhanced stars being \textit{brighter}. The derivatives are thus $dM_{V}/d\rm{[\alpha/\rm{Fe}]} =\; -$0.6 mag dex$^{-1}$ and $dM_{I}/d\rm{[\alpha/\rm{Fe}]} =\; -$0.3 mag dex$^{-1}$. 

We estimate the metallicity dispersion using the \citet{2008A&A...486..177Z} [Fe/H] measurements of the upper-RG branch. We find that between the two sightlines $b = -4^{\circ}$ and $b = -6^{\circ}$, the metallicity dispersion changes only slightly, dropping from 0.373 dex to 0.355 dex. Removing stars from the distribution with [Fe/H] $\leq -0.80$ -- RG stars that should eventually end up on the RR Lyrae instability strip and the horizontal branch but not the RC -- reduces the metallicity dispersion to 0.300 dex for $b = -4^{\circ}$ and 0.278 dex for $b = -6^{\circ}$. For [$\alpha$/Fe], we use the measurements of \citet{2010A&A...512A..41B}, averaging over [Ca/Fe], [Mg/Fe], and [Si/Fe]. Within the sample, ${\sigma}_{[\alpha/\rm{Fe}]}=$ 0.12 dex, and the value of the correlation coefficient is cor([$\alpha$/Fe],[Fe/H]) $= -$0.86.

Summing over these factors, we obtain:
\begin{eqnarray*}
\sigma_{I,\rm{Predicted}}^2 & = & \sigma_{I,\rm{single}}^2 + \sigma_{[\rm{Fe/H}]}^2 {\biggl[\frac{d\rm{M_{I}}}{d\rm{[Fe/H]}}\biggl]}^2 + \sigma_{[\alpha/\rm{Fe}]}^2{\biggl[\frac{d\rm{M_{I}}}{d[\alpha/\rm{Fe]}}\biggl]}^2 \\
& &  + 2
\rm{cor}([\alpha/\rm{Fe}],[\rm{Fe/H}])\sigma_{[\alpha/\rm{Fe}]}\sigma_{[\rm{Fe/H}]}
\biggl[\frac{\textit{d}M_{I}}{\textit{d}[\alpha/\rm{Fe]}}\frac{\textit{d}M_{I}}{\textit{d}\rm{[Fe/H]}}\biggl],
\end{eqnarray*}
yielding:
\begin{equation}
\sigma_{I,\rm{Predicted}} = 0.117 \rm{mag}
\end{equation}
A similar calculation for $\sigma_{V,\rm{Predicted}}$ yields a value of 0.256 mag, for a predicted intrinsic brightness dispersion ratio of $\sigma_{V}/\sigma_{I} \sim$ 2.18.

Our theoretical estimate is done under the assumption that the bulk of Galactic bulge stars  formed in a very small time-frame without enhanced helium-enrichment. The stars that are rich in [Fe/H] also have lower values of [$\alpha$/Fe] \citep{2010A&A...512A..41B,2011A&A...530A..54G}, as such a strong prediction of this single-age, simple-helium hypothesis is that the most metal-rich RC stars should on average be the dimmest. This trend is observed locally. \citet{2000ApJ...531L..25U}, in his analysis of Hipparcos RC stars with metallicity measurements, found that RC stars with [Fe/H]$\sim$0.0 were 0.08 mag fainter than RC stars with [Fe/H]$\sim-$0.36. If the most metal-rich stars are either substantially younger or more helium-enriched, then that correlation may be washed away or even reversed. 

We note the main reason why we obtain a larger estimate than \citet{2001MNRAS.323..109G}. It is primarily due to the the last term in Equation (1), the anti-correlation between [$\alpha$/Fe] and [Fe/H]. The stars with the highest [Fe/H] also have the lowest [$\alpha$/Fe], making them even fainter than expected relative to the stars with lower [Fe/H].

\section{Intrinsic Brightness Dispersion from Observations of the Double Clump}
\label{sec:DoubleClump}
The double RC of the Galactic bulge \citep{2010ApJ...721L..28N,2010ApJ...724.1491M}, in which two RCs with equal $(V-I)$ and $(J-K)$ color but with a 0.5 mag separation in brightness coexist, is a phenomenon observed toward sightlines $(|l| \lessapprox 3.0, |b| \gtrapprox 5.0)$. X-shaped bulges are ubiquitous in N-body models \citep{2005MNRAS.358.1477A} and in observations of field galaxies \citep{2006MNRAS.370..753B}. The two RCs show a significant mean difference in longitudinal proper motion \citep{2010ApJ...724.1491M} and are accompanied by two RGBBs of similar brightness and number counts \citep{2011ApJ...730..118N}. In light of these factors, we are confident in making the assumption that this is a geometric effect.

Studies of X-shaped bulges in other galaxies indicate that the geometrical dispersion toward these sightlines should be very small. The examples shown in Figure 1 of \citep{2006MNRAS.370..753B} have X-wings with total length typically $\sim$1/4 that of the bar's half-length. For a $\sim$2.5 kpc Milky War \citep{2007MNRAS.378.1064R}, that implies a total length of $\sim$600 pc. Dividing by $\sqrt{12}$ gives a maximum dispersion of 175 pc, or $\sim$0.05 mag for a structure at 8 kpc. We note that this is an upper-bound as the value of $\sqrt{12}$ is only correct if the stars are uniformly distributed within the X-wings, an unlikely propostion. If they are more heavily concentrated toward the center, the ratio of geometrical dispersion to total physical extent will be smaller, and the geometric dispersion consequentially reduced. 

\begin{figure}[H]
\begin{center}
\includegraphics[totalheight=0.5\textheight]{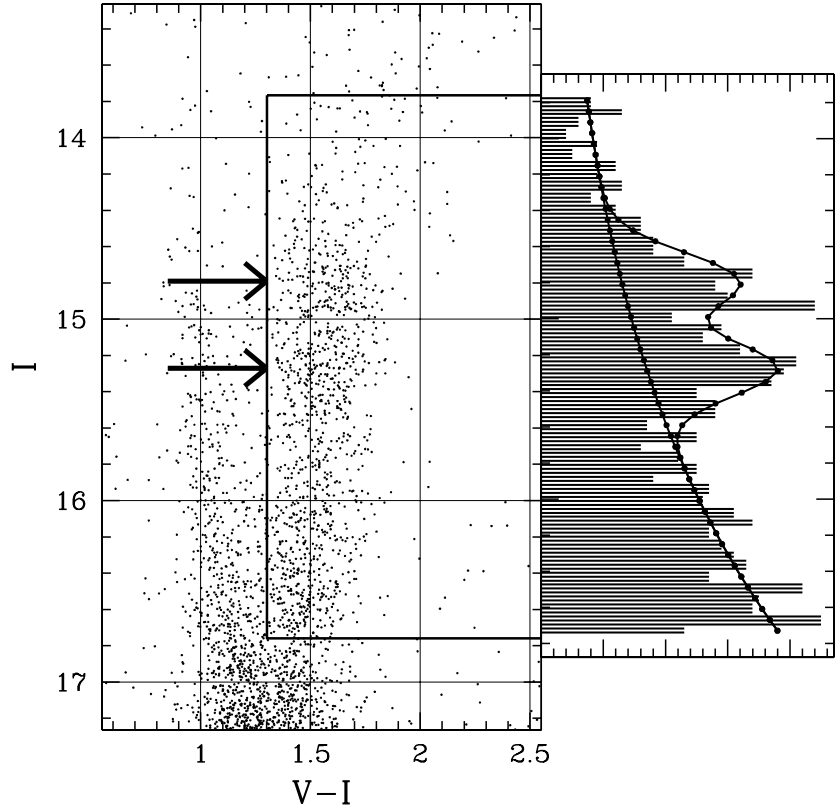}
\end{center}
\caption{The double RC toward the sightline (l,b) = (0.0, $-$6.3). The two peaks in the magnitude distribution of RG+RC stars are clearly discernible at $I=14.79$ mag and $I=15.27$ mag. The right panel shows the magnitude distribution of the RG+RC stars in $I$ for the subset of the stars in the left panel that are within the black shaded contour lines selecting the RG branch.} 
\label{Fig:DoubleClumpPresentationPlot2}
\end{figure}

We use a sample of 591 double RC measurements used as part of the study of  \citet{2010ApJ...721L..28N}. We fit for the RG + double RC stellar population using the equation:
\begin{equation}
N(m) = A\exp\biggl[B(m-m_{RC,1})\biggl] + \frac{N_{RC,1}}{\sqrt{2\pi}\sigma_{RC,1}}\exp\biggl[{-\frac{(m-m_{RC,1})^2}{2\sigma_{RC,1}^2}}\biggl] + \frac{N_{RC,2}}{\sqrt{2\pi}\sigma_{RC,2}}\exp\biggl[{-\frac{(m-m_{RC,2})^2}{2\sigma_{RC,2}^2}}\biggl]
\label{EQ:Exponential2}
\end{equation}
where the parameters are as described in \citet{2010ApJ...721L..28N}. For two Gaussians placed near each other, a fit will often have a reduced ${\chi}^2$ if the normalization and dispersion of one Gaussian is increased at the expense of that of the other Gaussian. We correct for this by imposing the geometrically-motivated prior that the two Gaussians should have the same brightness dispersion. We show a representative CMD in Figure \ref{Fig:DoubleClumpPresentationPlot2}.  

\begin{figure}[H]
\begin{center}
\includegraphics[totalheight=0.5\textheight]{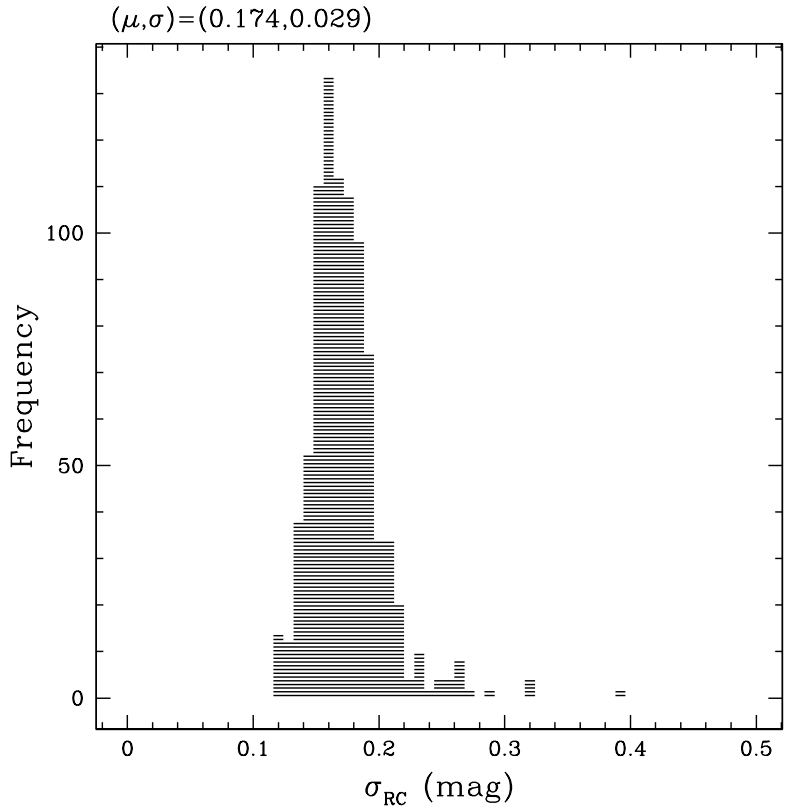}
\end{center}
\caption{Distribution of Gaussian widths ${\sigma}_{\rm{RC}}$ obtained for a sample of 591 sightlines toward the Milky Way's X-wings.} 
\label{Fig:DoubleClumpHist}
\end{figure}

We obtain a mean dispersion of 0.174 mag in $I$ for the two RCs. We plot the distribution of magnitude dispersions toward these sightlines in  Figure \ref{Fig:DoubleClumpHist}. We note that the standard deviation in the mean value of the dispersion obtained for these fields, 0.046 mag, is roughly equal to the average of the standard deviations for the width of the RCs in each of the Markov Chains used to obtain the best fit, 0.041 mag. This implies that all or nearly all of the variance seen in Figure \ref{Fig:DoubleClumpHist} is due to Poisson noise rather than structural changes.

Subtracting in quadrature the expected geometric dispersion of $\sim$0.05 mag from the obtained mean of 0.174 mag yields 0.167 mag. Therefore, our analysis of the double RC sightlines yields an intrinsic RC brightness dispersion ${\sigma}_{I} \approx 0.17$ mag.

\section{Measuring the Intrinsic Brightness Dispersion Using Clump-Centric Color Magnitude Diagrams in $V$ and $I$}
\label{sec:CCCMD}
\begin{figure}[H]
\begin{center}
\includegraphics[totalheight=0.6\textheight]{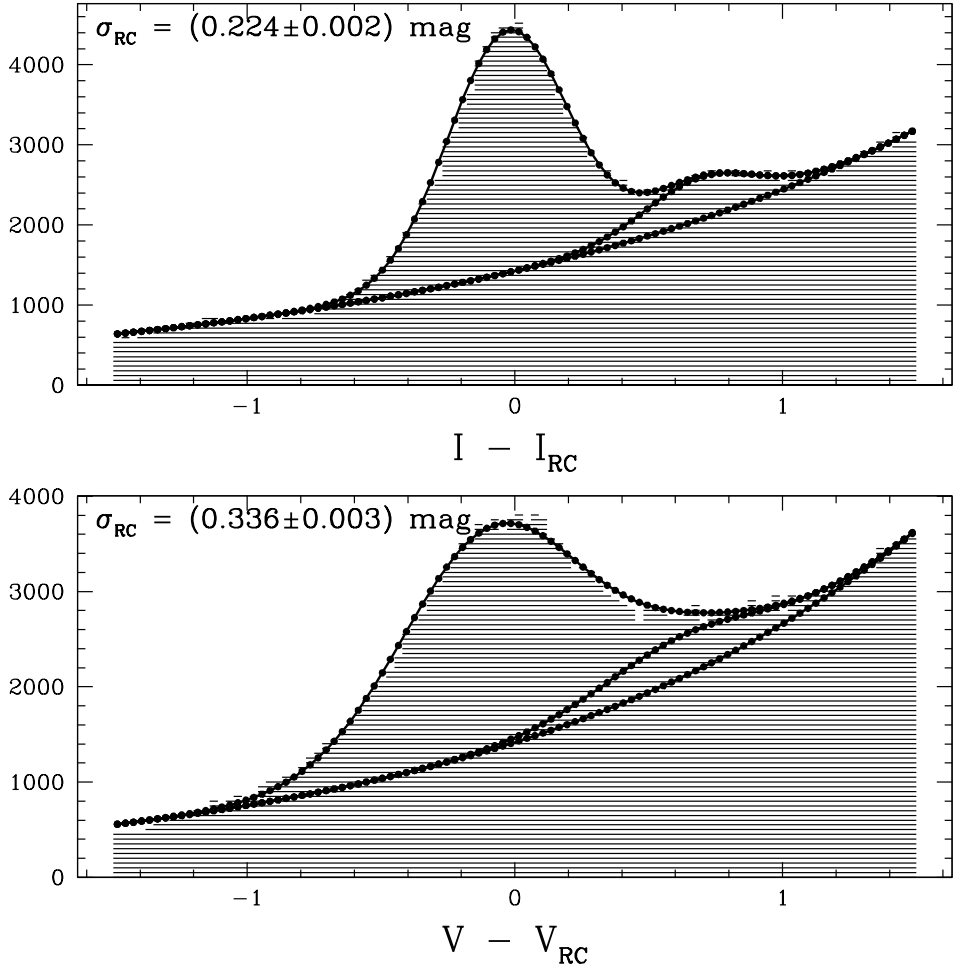}
\end{center}
\caption{TOP: Fit for the magnitude distribution in $I$ of $\sim$230,000 OGLE-III RG branch point sources near $(l,b) = (0,-2)$. BOTTOM: Same as top, but the magnitudes used are in $V$.} 
\label{Fig:DispersionPaperScript}
\end{figure}

We use CCCMDs to study the RC for OGLE-III point sources that are in OGLE-III subfields centered within 0.5 degrees of $(l,b) = (0,-2)$. The construction of the CCCMDs, in which the color and magnitude of each star is taken relative to that of the nearest measured RC centroid, is fully described in \citet{2011ApJ...730..118N}. Of the available sightlines in the OGLE-III bulge photometric catalog, $(l,b) = (0,-2)$ is among the ones that is closest to the plane, and will as such have the least amount of geometrically induced brightness dispersion for single-clump fields. The combined brightness+intrinsic dispersion is $\sim$0.22 mag, compared to $\sim$0.25 mag near Baade's window. This narrower dispersion allows a cleaner separation between the RC and the red giant branch bump (RGBB), reducing the degeneracy between the two. As in \citet{2011ApJ...730..118N}, we fit for the various components of the stellar population as follows:
\begin{equation}
N(m) = A\exp\biggl[B(I-I_{RC})\biggl] + \frac{N_{RC}}{\sqrt{2\pi}\sigma_{RC}}\exp \biggl[{-\frac{(I-I_{RC})^2}{2\sigma_{RC}^2}}\biggl]+\frac{N_{RGBB}}{\sqrt{2\pi}\sigma_{RGBB}}\exp \biggl[{-\frac{(I-I_{RGBB})^2}{2\sigma_{RGBB}^2}}\biggl]
\end{equation}
where $A$, $B$ are the parameters to the underlying RG branch, $N_{RC}$, $I_{RC}$, and ${\sigma}_{RC}$ are the normalization, mean brightness and brightness dispersion for the Gaussian used to model the RC; and $N_{RGBB}$, $I_{RGBB}$, and ${\sigma}_{RGBB}$ are the analogous parameters for the RGBB. About 230,000 sources are used to fit the combined  RC+RG+RGBB population. We obtain ${\sigma}_{RC} = (0.224 \pm 0.002)$ mag in $I$, with the quoted error being a statistical error. The resulting fit is shown in the top panel of Figure \ref{Fig:DispersionPaperScript}.

We repeat the process in $V$ though we constrain the RGBB parameters as they are less stable when the RC is broadened. The population ratio $f_{Bump} = \frac{N_{RGBB}}{N_{RC}}$ is fixed to have the same value as obtained when fitting in $I$, the magnitude difference is fixed to be the same (an assumption of identical mean $(V-I)$ color for the RGBB and RC), and ${\sigma}_{RGBB}$ is fixed to 0.3 mag. The resulting fit is shown in the bottom panel of Figure \ref{Fig:DispersionPaperScript}. We obtain ${\sigma}_{V} = (0.336 \pm 0.003)$ mag. We note that the value obtained here is very weakly sensitive to the reasonable priors we have imposed on the RGBB in $V$. Removing the constraint that the RGBB and RC to have the same mean color raises $\sigma_{V}$ from 0.336 to 0.343 mag, whereas floating ${\sigma}_{RGBB}$ as a free parameter in $V$ raises $\sigma_{V}$ from 0.336 to 0.347 mag. 

A few results are immediately apparent from Figure \ref{Fig:DispersionPaperScript}. First, the RC is much more well-defined in $I$. There is no genuine prospect, for example, of ever measuring the parameters of the Galactic bulge RGBB in $V$. The secondary overdensity in simply not discernible in $V$, and its parameters must be imposed. 

We comment on the validity of the assumption that the brightness of the RC is distributed as a Gaussian. It is known from globular cluster studies that the RC of a single-age, single-metallicity, single-distance population is intrinsically non-Gaussian -- there's a steep rise in number counts at the location of the zero-age horizontal branch, followed by a slow gradual decline. However, these sightlines have additional noise due to the broad metallicity variation and geometrical thickness of the Galactic bulge. Indeed, the simulated bulge RCs in Figure 14 of  \citet{2001MNRAS.323..109G} have very Gaussian-looking brightness distributions, and that is without the additional effect of geometrical broadening. 

The intrinsic magnitude dispersion  $\sigma_{I}$ can be estimated from the combined geometric and intrinsic magnitude dispersion in $I$ and $V$. We assume that the RC in $I$ and $V$ will have the same contribution from geometric dispersion, but different contributions due to population effects:
\begin{equation}
\sigma_{RC,I}^2 = \sigma_{G}^2 + \sigma_{I}^2
\end{equation}
\begin{equation}
\sigma_{RC,V}^2 = \sigma_{G}^2 + \sigma_{V}^2,
\end{equation}
Where $\sigma_{G}$ is the geometric dispersion, $\sigma_{I}$ is the intrinsic dispersion in $I$, and analogously for $\sigma_{V}$. Substituting in $\sigma_{I}=0.224$ and $\sigma_{V}=0.336$ and taking the difference of the equations yields:
\begin{equation}
\sigma_{V}^2 - \sigma_{I}^2 = 0.0672
\end{equation}
The value of $\sigma_{I}$ is then a monotonic function of the ratio $\sigma_{V}/\sigma_{I}$. 

Adopting the expected value of $\sigma_{V}/\sigma_{I} = 2.18$ derived in Section \ref{sec:sources} yields $\sigma_{I} = 0.14$ mag. A consideration of the uncertainties in this estimate -- and thus of the uncertainty in $\sigma_{V}/\sigma_{I}$ -- is in order. Of the effects listed, the only one that in and of itself produces a ratio less than  $\sigma_{V}/\sigma_{I} = 2.00$ (for which $\sigma_{I} = 0.15$) is that of the luminosity evolution of a RC star during its lifetime. It is $\sim$0.06 mag for 47 Tuc. If the value were \textit{doubled}, to 0.12 mag, the ratio would be lowered to $\sim$1.8, yielding $\sigma_{I} = 0.17$. Alternatively, using the value of $\sigma_{I} = 0.17$ measured in Section \ref{sec:DoubleClump} yields $\sigma_{V}/\sigma_{I} = 1.8$.

We therefore adopt $\sigma_{I} \approx$ 0.14-0.17 mag as the estimate obtained by comparing the dispersions in $V$ and $I$.

\subsection{On the Assumption of Minimal Differential Reddening}
At the request of the anonymous referee, we discuss the assumption of minimal differential reddening within the sightlines used to construct the CCCMD used in section 5. 

We first note that the sightlines used to measure the RC centroids before building the CCCMD are no more than 3$\arcmin$ apart, which means that any point source kept will be within $\sim$2$\arcmin$ of a measured RC centroid. This maximizes any differential reddening to that leftover on $\sim$2$\arcmin$ scales. Second, as discussed in \citet{2011ApJ...730..118N},  when constructing our clump-centric photometry files we directly visually inspected each sightline and discarded those with very high differential reddening, for example those where the RC would have a slanted profile on the CMD.

\begin{figure}[H]
\begin{center}
\includegraphics[totalheight=0.4\textheight]{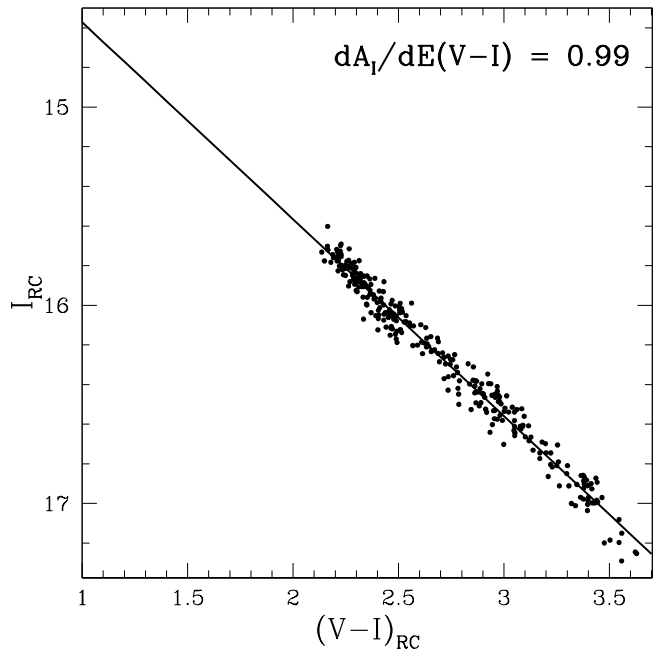}
\end{center}
\caption{The color and magnitude of 304 RC centroids measured within 0.5 degrees of  $(l,b) = (0,-2)$.} 
\label{Fig:DiffReddening}
\end{figure}

The 304 RC centroids measured within 0.5 degrees of the sightline $(l,b) = (0,-2)$ have a standard deviation in coordinate space of $\sim$15$\arcmin$, and a standard deviation in reddening of 0.4 mag in $E(V-I)$. We show the scatter plot in Figure \ref{Fig:DiffReddening}. This implies that even if the most egregious cases of differential reddening were not directly removed, the expected differential reddening per sightline would be $\sim$0.05 mag in E(V-I) and thus the same in $I$ due to the reddening law toward these sightlines, $A_{I}/E(V-I)$ = 0.99. As the intrinsic dispersion estimated in Section 5 is  $\sigma_{I} \approx$ 0.14-0.17 mag, differential reddening will not significantly contribute in quadrature.

\section{Conclusion}
\label{sec:Conclusion}
In this Letter, we obtain similar values for the intrinsic brightness dispersion of the Galactic bulge RC using two different methods, by analyzing the fits to double RCs toward $(-3.5 \lessapprox l \lessapprox 1.5, |b| \gtrapprox 5.0)$, and by comparing CCCMDs toward $(l,b) \approx (0,-2)$ . We respectively obtain values of ${\sigma}_{I} \approx 0.17$ and ${\sigma}_{I} \approx 0.14-0.17$ mag using the two different methods. 
Our result has consequences for studies of Galactic structure. \citet{1997ApJ...477..163S} found that decreased values of the intrinsic magnitude dispersion of the RC would yield decreased values of the bar angle  $\alpha$ and the bar structural parameter $x_{0}$:$z_{0}$.

These value ${\sigma}_{I}\approx 0.17$ mag is larger than the theoretical estimates of $\sim$0.051-0.107 mag found in \citet{2001MNRAS.323..109G}. It is also higher than our theoretical estimate of $\sim$0.117 mag. The difference may be due to errors in the theory, binary evolution effects, or correlated variations between the input parameters we considered with age and helium-enrichment. 

\acknowledgments
DMN was partially supported by the NSF grant AST-0757888. The OGLE project has received funding from the European Research Council
under the European Community's Seventh Framework Programme
(FP7/2007-2013) / ERC grant agreement no. 246678 to AU.

\end{document}